\documentclass{article}

    \PassOptionsToPackage{numbers}{natbib}
     \usepackage[preprint]{neurips_2021}

\usepackage[utf8]{inputenc} % allow utf-8 input
\usepackage[T1]{fontenc}    % use 8-bit T1 fonts
\usepackage{hyperref}       % hyperlinks
\usepackage{url}            % simple URL typesetting
\usepackage{booktabs}       % professional-quality tables
\usepackage{amsfonts}       % blackboard math symbols
\usepackage{nicefrac}       % compact symbols for 1/2, etc.
\usepackage{microtype}      % microtypography
\usepackage{xcolor}         % colors
\usepackage{graphicx}
\usepackage{soul}
\usepackage{comment}

\title{How Mock Model Training Enhances\\ User Perceptions of AI Systems}

\author{
  Amama Mahmood, Gopika Ajaykumar, and Chien-Ming Huang\\
  \texttt{\{amama.mahmood, gopika.ajaykumar, chienming.huang\}@jhu.edu}\\
  Department of Computer Science\\
  Johns Hopkins University\\
  Baltimore, MD 21218, USA \\
 }

\begin{document}

\maketitle

\begin{abstract}
Artificial Intelligence (AI) is an integral part of our daily technology use and will likely be a critical component of emerging technologies. However, negative user preconceptions may hinder adoption of AI-based decision making. Prior work has highlighted the potential of factors such as transparency and explainability in improving user perceptions of AI. We further contribute to work on improving user perceptions of AI by demonstrating that bringing the user in the loop through mock model training can improve their perceptions of an AI agent's capability and their comfort with the possibility of using technology employing the AI agent.

\end{abstract}
\section{Motivation and Background}

Fueled by increasingly available user data, growing computing power, and recent advances in machine learning, Artificial Intelligence (AI) technologies are transforming our society and daily lives. However, users' negative preconceptions of AI may hinder adoption and continued use of AI technologies. Negative user preconceptions can affect user trust, which is a key factor in determining acceptance of technology \citep{venkatesh2003user}. Inadequate user trust can in turn lead to misuse (i.e., inappropriate reliance on technology) and disuse (i.e., underutilization of technology due to rejection of its capability) \citep{parasuraman1997humans,lee2004trust}. 
To enhance user perceptions of AI systems, previous research has investigated AI transparency, explainability, and interpretability (e.g., \citep{adadi2018peeking, arrieta2020explainable}), as modern machine learning methods are largely black boxes \citep{holzinger2018current, castelvecchi2016can}. %check: castelvecchi2016can 
For example, prior work has explored how visualization may aid user understanding of how machine learning models work (e.g., \citep{samek2019explainable, choo2018visual}). 
Explanations of these models and justifications for decisions made by intelligent machines help users understand their inner workings once they begin interacting with the AI technologies. In this work, we explore how to improve users' existing \emph{preconceptions} of AI agents prior to any interactions with the agents.

Simulated setups, such as mock trials, mock interviews, and drills, have been used as low-cost, hands-on tools in early training phases to help people become accustomed to unfamiliar practices and processes prior to engaging in them. Similarly, we explore the potential of using mock interactions in which users label training data for AI models in modulating users' confidence in AI agents' capabilities and their comfort with the possibility of using technologies employing the AI agents before engaging in real interactions with the AI agents. We contextualize our exploration within the scenario of training AI agents for use in autonomous vehicles---a safety-critical domain that is likely to involve interactions with everyday users. Our findings indicate that users' perceptions of AI agents improved through participation in mock model training, especially when they were able to precisely label objects that they perceived to be important.

\begin{figure*}
    \centering
    \includegraphics[width=1.0\textwidth]{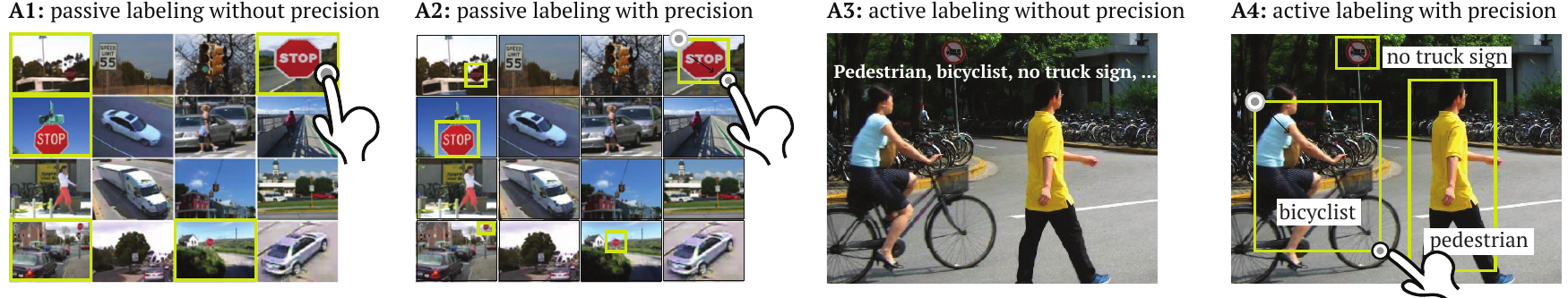}
    \caption{We explore how mock model training involving various data labeling strategies may affect users' perceptions of AI agents posed as driving assistants.}
    \label{fig:exp-design}
\end{figure*}
\section{Methods}
\label{sec:methods}
\subsection{Experimental Design, Task, and Conditions}
\label{sec:setup}
We conducted a within-subjects study that consisted of four experimental conditions (Figure \ref{fig:exp-design}). %and a baseline as reference.
The study was contextualized within the scenario of labeling images to train four AI agents to perform driving-related object identification:
\begin{itemize}\itemsep0em 
\item \textbf{A1}: To train this agent, the participant was presented with a grid of images that included five positive examples for each of six item categories commonly encountered during driving: stop sign, speed limit sign, traffic light, car, bicyclist, and pedestrian. %16 images
This labeling process is similar to image selection tasks commonly used in web security checks. 
It represents low labeling precision (i.e., the user did not localize the object within the image) and passive labeling (i.e., the user only labeled items for the requested categories). It is analogous to binary object \emph{detection} (i.e., indicating whether or not a specified item is present).

\item \textbf{A2}: The participant followed a similar labeling process to train agent A2 as done for A1, with the additional task of drawing bounding boxes around the target item in all images. 
This process represents high labeling precision and is analogous to binary object \emph{recognition}..  

\item \textbf{A3}: In training this agent, the participant was provided with a set of individual images for labeling. For each image, the user was prompted to list all items within the image that they considered to be relevant via text. The user was free to specify as many item categories as they wanted. This method is analogous to multiple object \emph{detection}.%12 images

\item \textbf{A4}: Similarly to the training task for A3, the participant was prompted to draw bounding boxes around all items that they considered to be relevant and to specify the associated labels via text within each image in the set.
This process represents high labeling precision and is analogous to multiple object \emph{recognition}.
\end{itemize}

We also presented a baseline pre-trained agent to the participant at the beginning of the study. The participant was able to review the images used to the train the agent. We used this baseline condition as a reference to measure users' preconceptions of an AI agent without undergoing mock training.

\subsection{Measures}
We used a range of metrics to measure user perceptions that may affect user trust in and adoption of AI technologies. For each trained agent, we computed the \emph{difference} in comfort, projected capability, and task confidence relative to the baseline, pre-trained agent (i.e., positive values indicate an improvement in user perceptions relative to the baseline). We normalized the data from all questionnaire responses to get values in a 0 -- 1 range before computing the difference.
\begin{itemize} \itemsep0em 
    \item \textbf{Trustworthiness.} Trust was measured through a single question asking which AI agent the participant would trust the most if it was employed in an autonomous vehicle.
    \item \textbf{Comfort.} Comfort was measured through a custom scale %(Scale: 1--6) 
   consisting of six statements (Cronbach's $\alpha=0.90$) prompting users to rate how comfortable they felt towards a self-driving car employing the trained agent (Appendix \ref{app:comfort}).

    \item \textbf{Projected Capability.} Projected capability was measured through a custom scale consisting of four statements (Cronbach's $\alpha=0.87$) prompting participants to rate  how capable they felt the self-driving car employing the trained agent to be (Appendix \ref{app:cabability}).

    \item \textbf{Task Confidence.} To quantify their perception of the AI agent's performance, we asked participants to rate their confidence (0--100\%) in its ability to identify specified items (e.g., stop sign) for a set of 14 images. This set included two images for each of the six item categories (12 images) and two images representing ``unseen'' items (e.g., no-left-turn sign and pedestrian-crossing sign) that were not included in the object categories used for training agents A1 and A2.
\end{itemize}

\subsection{Procedure}
\label{sec:procedure}
The study consisted of five phases: 
(1) \textit{Introduction and consent}. 
Upon opening the website, participants were briefed about the study and were informed that they would be training AI agents to become driving assistants by providing examples of things (e.g., stop signs and pedestrians) that the agents may encounter on the road.
(2) \textit{Reference}. 
The participants review the images used to train the baseline, pre-trained agent and complete the confidence assessment and perception survey.
(3) \textit{Labeling training examples for AI agents A1-A4}. 
The participants labeled training data for for the four experimental conditions, which were counterbalanced using a Latin square design.
(4) \textit{Confidence assessment and perception survey}. 
Participants were asked to rate task confidence and questions about trust. They then continued to the next condition and repeated phase 3--4.
(5) \textit{Post-study questionnaire}. 
At the end, participants filled out a post-study questionnaire, which asked which agent they trusted the most and collected demographics information. 
The study was approved by our institutional review board and took approximately 45 minutes to complete. The participants were compensated with \$10 USD upon completion of the study.

\begin{figure*}
    \centering
    \includegraphics[width=1.0\textwidth]{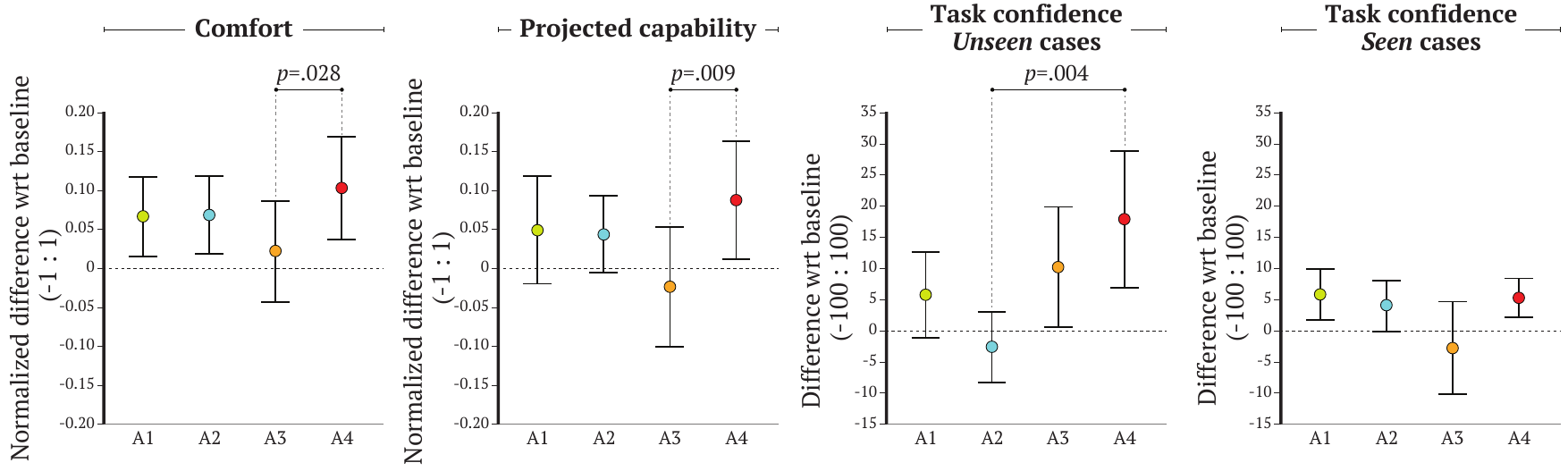}
    \caption{One-way repeated measures ANOVAs were conducted to discover effects of experimental condition on comfort, projected capability, and task confidence for seen and unseen cases. Error bars represent 95\% confidence intervals; only significant comparisons ($p<.05$) are highlighted.}
    \label{fig:results}
\end{figure*}

\section{Results}
A total of 35 participants (17 females, 17 males, 1 non-binary) were recruited for this online study via convenience sampling. The participants were aged between 18 to 35 ($M=25.91, SD=4.78$) and were from a variety of educational backgrounds, including computer science, engineering and technology, social work, healthcare, life sciences, business, law, media, public policy, and education. 
The participants reported having minimal experience with self-driving cars ($M=1.40, SD=0.85$), and moderate experience with AI products ($M=3.49, SD=1.72$) and with training AI or machine learning models ($M=3.09, SD=1.69$), using 6-point rating scales with 1 being no experience and 6 being lots of experience.
Figure \ref{fig:results} summarizes our main findings. 
For all statistical tests reported below, $p<.05$ was considered a significant effect.
We followed Cohen's guidelines on effect size and considered $\eta_p^2=0.01$ a small effect size, $\eta_p^2=0.06$ a medium effect size, and $\eta_p^2=0.14$ a large effect size \citep{cohen1988statistical}. 

A chi-square goodness-of-fit test showed that users did not perceive AI agents, including the baseline agent, as equally trustworthy, $\chi^2(4,35)=43.60, p<.001,  v=0.56$. In particular, A4 (active labeling with high precision) was considered the most trustworthy agent by the most participants (51\%). 
A one-way repeated measures analysis of variance (ANOVA) yielded a significant main effect of experimental condition on comfort, $F(3,102) = 3.75, p=.013, \eta_{p}^{2} = .099$. Post-hoc pairwise comparisons with a Bonferroni correction revealed that comfort increased with active labeling with precision, A4 ($M=0.10, SD=0.19$), more than  with active labeling without precision, A3 ($M=0.02, SD=0.22$), $p=.028$. 
Moreover, a one-way repeated measures ANOVA yielded a significant main effect of the experimental condition on projected capability, $F(3,102) = 4.69, p=.004, \eta_{p}^{2} = .121$. Post-hoc pairwise comparison with a Bonferroni correction revealed that active labeling with precision, A4 ($M=0.09, SD=0.22$), had higher improvement in projected capability than active labeling without precision, A3  ($M=-0.02, SD=0.22$), $p=.009$. 

A one-way repeated measures ANOVA yielded a significant main effect of the experimental condition on task confidence for \textit{unseen} cases, $F(3,102) = 6.76, p<.001, \eta_{p}^{2} = .166$. Post-hoc pairwise comparisons with a Bonferroni adjustment revealed that active labeling with precision, A4 ($M=18.00, SD=32.16$), had higher improvement in task confidence than passive labeling with precision, A2 ($M=-2.57, SD=16.47$), $p=.004$. 
While a one-way repeated measures ANOVA yielded a significant main effect of the experimental condition on task confidence for seen cases, $F(3,102) = 4.44, p=.006, \eta_{p}^{2} = .115$, we did not observe any significant differences in pairwise comparisons.

\section{Discussion}
In this study, we observed that users associated higher levels of comfort and projected capability with the agents for which they labeled training data with precision. Moreover, for unseen cases, users perceived the agent for which they were able to freely label objects of interest to be more capable. Our results suggest that everyday users can perceive the importance of high-precision training data representative of diverse scenarios in determining AI task performance. Therefore, involving users in mock training exercises where they obtain hands-on experience with training data may help them in developing accurate mental models of how an AI agent operates and in maintaining appropriate trust levels in the AI agent's performance before working with or using the AI technology. Furthermore, our study suggests that greater levels of user involvement (e.g., precise labeling using bounding boxes) may help users feel more comfortable with using an AI agent, even in a more safety-critical scenario. Overall, our study suggests that mock training setups can serve to help set up appropriate user understanding and improved preconceptions of how an AI agent will operate prior to real interaction with the AI agent.

\label{sec:limitations}
One of the limitations of this study is that we focused on user trust in AI through a questionnaire item, rather than relying on behavioral (e.g., \cite{yu2019trust}) or physiological (e.g., \cite{hergeth2016keep}) measures. As a result, we may have failed to accurately or fully capture actual user trust in AI systems.
In future studies, we would like to investigate alternate methods for measuring and investigating trust so that we can better understand the range of factors that contribute to user trust in human-AI interaction. 
We would also like to expand our study of mock model training in AI systems to encompass new types of interactions in different domains. In this work, we chose to contextualize our study within the scenario of training AI agents for self-driving cars, which is a safety-critical domain that many users may not have direct experience with. Therefore, we would like to further investigate how our findings would apply to more general, commonplace scenarios that may involve lower stakes, such as speech-based interactions with AI agents in smart-speakers. 
Furthermore, we investigated the effects of mock model training as \textit{explicit} participation in this work, but users may participate in different forms and in other phases of machine learning, such as algorithm design or error correction. Furthermore, modern machine learning systems may involve users without their knowledge or explicit consent, such as recommender systems used in online services. Future work should investigate if user participation still positively influences perceptions of AI in cases where users are engaged outside of AI training or implicitly without their awareness. 

\section*{Acknowledgements}
This work was supported by the National Science Foundation award
\#1840088, the National Science Foundation Graduate Research Fellowship Program under Grant No.
DGE-1746891, and the Nursing/Engineering joint fellowship from the Johns Hopkins University. 

\newpage
\bibliographystyle{unsrtnat}
\bibliography{references}

%%%%%%%%%%%%%%%%%%%%%%%%%%%%%%%%%%%%%%%%%%%%%%%%%%%%%%%%%%%%

%\newpage
\appendix

\section{Appendix}

\subsection{Comfort -- Cronbach's $\alpha=0.90$}
\label{app:comfort}
Please rate the following regarding the self-driving car that has employed Driving Assistant X: 
\begin{itemize}
    \item I would be wary of the self-driving car\footnotemark
    \item I would be afraid that the self-driving car would be harmful\footnotemark[\value{footnote}]
    \item I would be confident riding in self-driving car
    \item I would be comfortable riding in the self-driving car
    \item I would be relaxed while riding in the self-driving car
    \item I would be agitated while riding in the self-driving car\footnotemark[\value{footnote}]

\end{itemize}

\subsection{Projected capability -- Cronbach's $\alpha=0.87$}
\label{app:cabability}
Please rate following regarding the self-driving car that has employed Driving Assistant X:
\begin{itemize}
    \item I believe that the self-driving car would NOT be dependable\footnotemark[\value{footnote}]
    \item I believe that the self-driving car would be reliable
    \item I would trust the self-driving car to identify pedestrians, signs and signals, and obstacles correctly
    \item I am confident that the self-driving car would comply with traffic rules
\end{itemize}
\footnotetext{Reverse scale items}
\subsection{Image Sources}
\label{app:sources}
The images that we used in the study included public domain images from the web and images from various datasets, including the \textit{Penn-Fudan Database for Pedestrian Detection and Segmentation} \cite{wang2007object}, the \textit{MIO-TCD Dataset} \cite{luo2018mio}, the \textit{LISA Traffic Sign Dataset} \cite{mogelmose2012vision}, and \textit{ImageNet} \cite{deng2009imagenet}. 
Figure \ref{fig:exp-design} shows examples of images used for our study tasks.

\end{document}